# Highlights

**Advanced-Glycation Endproducts: How cross-linking properties affect the collagen fibril behavior**

Julia Kamml,Claire Acevedo,David S. Kammer

- High stiffness, critical length and contents of cross-links cause stiffening of the collagen fibril at high strain levels.

- Peak stress of collagen fibril depends on AGEs density

- The stiffening of the fibril is caused by reduced sliding between the tropocollagen molecules due increased energy capacity of cross-links and AGEs density

- When inter-molecular sliding is reduced, stretching of the tropocollagen molecules increases

- Decreased sliding leads to less energy dissipation - increased stretching leads to energy absorption in the tropocollagen molecules

- When energy is absorbed via stretching rather then sliding, fibrillar failure is shown to be brittle.

# Advanced-Glycation Endproducts: How cross-linking properties affect the collagen fibril behavior

Julia **Kamml**[a], Claire **Acevedo**[b,c] and David S. **Kammer**[a,*]

[a]*Institute for Building Materials, ETH Zurich, Switzerland*
[b]*Department of Mechanical Engineering, University of Utah, Salt Lake City, Utah, USA*
[c]*Department of Biomedical Engineering, University of Utah, Salt Lake City, Utah, USA*

## ABSTRACT

Advanced-Glycation-Endproducts (AGEs) are known to be a major cause of impaired tissue material properties. In collagen fibrils, which constitute a major building component of human tissue, these AGEs appear as fibrillar cross-links. It has been shown that when AGEs accumulate in collagen fibrils, a process often caused by diabetes and aging, the mechanical properties of the collagen fibril are altered. However, current knowledge about the mechanical properties of different types of AGEs, and their quantity in collagen fibrils is limited owing to the scarcity of available experimental data. Consequently, the precise relationship between the nano-scale cross-link properties, which differ from type to type, their density in collagen fibrils, and the mechanical properties of the collagen fibrils at larger scales remains poorly understood. In our study, we use coarse-grained molecular dynamics simulations and perform destructive tensile tests on collagen fibrils to evaluate the effect of different cross-link densities and their mechanical properties on collagen fibril deformation and fracture behavior. We observe that the collagen fibril stiffens at high strain levels when either the AGEs density or the loading energy capacity of AGEs are increased. Based on our results, we demonstrate that this stiffening is caused by a mechanism that favors energy absorption via stretching rather than inter-molecular sliding. Hence, in these cross-linked collagen fibrils, the absorbed energy is stored rather than dissipated through friction, resulting in brittle fracture upon fibrillar failure. Further, by varying multiple AGEs nano-scale parameters, we show that the AGEs loading energy capacity is, aside from their density in the fibril, the unique factor determining the effect of different types of AGEs on the mechanical behavior of collagen fibrils. Our results show that knowing AGEs properties is crucial for a better understanding of the nano-scale origin of impaired tissue behavior. We further suggest that future experimental investigations should focus on the quantification of the loading energy capacity of AGEs as a key property for their influence on collagen fibrils.

## 1. Introduction

The accumulation of Advanced-Glycation Endproducts (AGEs) in the body is a major concern for elderly individuals and patients with diabetes (Peppa and Vlassara, 2005). It has been demonstrated that AGEs are responsible for a large number of negative effects, such as chronic kidney diseases or an increased risk of macrovascular and microvascular complications (Rabbani and Thornalley, 2018; Di Pino et al., 2017; Hegab et al., 2012). The presence of AGEs is also associated with impaired material properties in collagenous tissues, such as tendons, vertebrae, bones, or the cornea (Li et al., 2013; Vashishth, 2009; Furst et al., 2016; Sato et al., 2001). For example, in bone, an increased AGEs content has been observed to be correlated with an increased fracture risk and bone brittleness (Rubin et al., 2016; Hunt et al., 2019; Vashishth et al., 2001; Tang et al., 2007, 2009; Acevedo et al., 2018a). Based on this correlation, it is commonly assumed that AGEs are an important source of reduced functionality in elderly and diabetic patients because these groups show increased AGEs content in their tissues (Saito et al., 1997). However, the precise mechanisms by which AGEs affect tissue behavior and impair body function remain unknown.

Collagen fibrils, which constitute a fundamental component of tissue at the nanoscale, are highly susceptible to the accumulation of AGEs. Among the various types of collagen, collagen type I is the most abundant in the human body. Its fibrils exhibit a distinctive banding pattern, which is the result of the five-staggered assembly of tropocollagen (TC) molecules, each comprising three polypeptide helices (see Fig. 1a,b). The non-enzymatic glycation process, known as the Maillard reaction, leads to the formation of AGEs at lysine residues of the TC molecules in the presence of sugars. Previous studies have demonstrated that the accumulation of AGEs impacts the mechanical properties of collagen fibrils (Andriotis et al., 2019; Gautieri et al., 2017; Kamml et al., 2023). However, the specific mechanisms caused by these changes within the fibril structure remain poorly understood. Since the macroscale behavior of the tissue is directly influenced by the behavior of collagen fibrils, understanding the effects of AGEs at the nanoscale is crucial for uncovering alterations in tissue behavior caused by AGEs accumulation.

The correlation of AGEs density and altered collagen behavior has been shown in several studies (Fessel et al., 2014; Li et al., 2013; Andriotis et al., 2019; Gautieri et al., 2017), but AGEs occur in different types and functions (Schmitt et al., 2005; Karim and Bouxsein, 2016; Perrone et al., 2020; Sell et al., 2005). Concerning their functions, we distinguish between cross-linking and non-cross-linking

*Corresponding author
✉ dkammer@ethz.ch (D.S. Kammer)
orcid(s): 0000-0003-3782-9368 (D.S. Kammer)

AGEs, where AGEs can either accumulate intramolecular bonding to a single TC-molecule or intermolecular by cross-linking two or more TC-molecules within the collagen fibril. Both non-crosslinking and crosslinking AGEs affect the collagen properties but in different ways. The non-crosslinking AGEs may cause changes in the tissue metabolism and protein function, whereas the cross-linking AGEs affect the mechanical behavior of the collagen fibril at the nanoscale, which may lead to changes at the tissue level. A computational study has shown that glucosepane, the most abundant AGE in tissue, has 8 potential cross-linking binding sites (Gautieri et al., 2014), and may cause fibrillar stiffening (Kamml et al., 2023), but otherwise, there is very limited information about their exact quantity and location, since common imaging techniques do not provide access to the nanoscale within the collagen fibril. Hence, it remains unknown whether AGEs cross-linking is causative to the observed impairment of mechanical properties in the tissue, as the exact relations between cross-link functions, mechanical properties, density and material behavior has not been determined yet (Willett et al., 2022).

Several types of AGEs cross-links have been identified so far, *e.g.*, glucosepane, pentosidine, GLUCOLD, crossline, vesperlysine, GOLD and MOLD (Svensson et al., 2018; Saito et al., 1997, 2011; Nemet et al., 2011; Nakamura et al., 1992; Yamaguchi et al., 1998; Vistoli et al., 2013). Each of these AGEs presents a different structural configuration and, therefore, has different mechanical properties, *e.g.*, strength and stiffness. While various experimental studies have shown the influence of AGEs content on the mechanical properties of collagenous tissues (Acevedo et al., 2018b; Yang et al., 2012; Svensson et al., 2013; Zellers et al., 2021), it is unclear which type of AGEs is responsible for these observations. Specifically, there is no complete information about the content of the various types of AGEs in these experiments, and hence, current studies cannot link the mechanical property of the collagen fibril to a specific AGE or a specific (mechanical) AGE property. In other words, knowledge about how AGEs cross-link properties translate to the fibril behavior remains missing, which prevents quantitative prediction of acceptable AGEs content *per type* for healthy collagen fibril mechanics.

In this paper, we study the link between the mechanical properties of AGEs cross-links at the nanoscale and the properties of the collagen fibril at the larger scale to evaluate how different types of AGEs cross-links influence the collagen behavior. We perform destructive tensile tests on a 3D coarse-grained steered molecular dynamics model of the collagen fibril with randomly inserted cross-linking in the helical regions of the TC molecules mimicking the contribution of AGEs. The properties of the AGEs cross-links, including stiffness and critical fracture length are varied systematically to account for differences in the structure of different AGEs cross-links. We quantify their effect on the mechanical properties of the collagen fibril at various cross-link densities with particular focus on fibril stiffness and strength, but most importantly, we show their influence on the deformation and fracture mechanism of the fibril on the molecule level.

## 2. Material and methods

We build a 3D coarse grained model of a representative collagen fibril and perform tensile tests using steered molecular dynamics. AGEs cross-links are inserted randomly along the TC-molecules at different densities and we vary cross-link parameters, namely stiffness and critical bond breaking length. The model is based on our previous study on the influence of cross-link density on the mechanical properties of the collagen fibril (Kamml et al., 2023). Our fibril geometry was implemented without enzymatic cross-links, since we could show that additional enzymatic cross-links do not influence or modify fibril mechanics in a crucial way. In our approach, TC molecules are represented by particles arranged in a chain and we apply particle interactions according to multi-body potentials. The chains are placed in a staggered fashion to form a collagen fibril that is consistent with their physiological arrangement (see Fig. 1a-d)(Buehler, 2006a,b; Depalle et al., 2015; Kamml et al., 2023). For a consistent overview of the applied parameters in the coarse-grained model, please refer to Tab. 1 and Appendix 8.1.

### 2.1. Geometry of the collagen fibril and insertion of cross-links

The geometry of the collagen fibril model is built to represent the biological configuration of collagen type I in tissue. Specifically, the TC-molecules are arranged in a 5-staggering pattern with gap $(0.6 \cdot D)$ and overlap $(0.4 \cdot D)$ zones and a periodicity of $D = 67\,nm$, also represented in Fig. 1a-d. One period with 5 gap- and overlap zones is used as a representative part of the collagen fibril (see Fig. 1c). The TC-molecules are represented by chains of particles where the particle-interaction potential models the mechanical behavior of the molecules (see Fig. 1b). Each of these particle chains has a length of $300\,nm$ and a diameter of about $1.5\,nm$, represented by the dispersive parameter $\sigma = 1.472\,nm$. The geometry of the single TC-molecules was obtained from Protein Data Bank entry 3HR2, the atomistic model based on X-ray crystallography (Orgel et al., 2006). In total, 218 particles are placed equidistantly $(r_0 = 14.0\,\text{Å})$ along a spline which was fitted along the longitudinal direction of the TC-molecule. In nature, collagen fibrils are bundles of TC-molecules with a diameter between 20 and several hundred nanometers (Hulmes, 2008). The modeled representative part of the collagen fibril has a diameter $d = 20.2\,nm$ and includes 155 TC molecules aligned through the cross section along the longitudinal axis. At the ends of the collagen fibril specimen, where forces are applied, the TC-molecules are extended with 40 additional particles with strengthened bonds to guarantee smooth force transition.

AGEs cross-links, which are at the focus of our study, are naturally built via glycation between the centrally located helical regions of the TC-molecules. In our model, we insert AGEs cross-links randomly between the central 95%

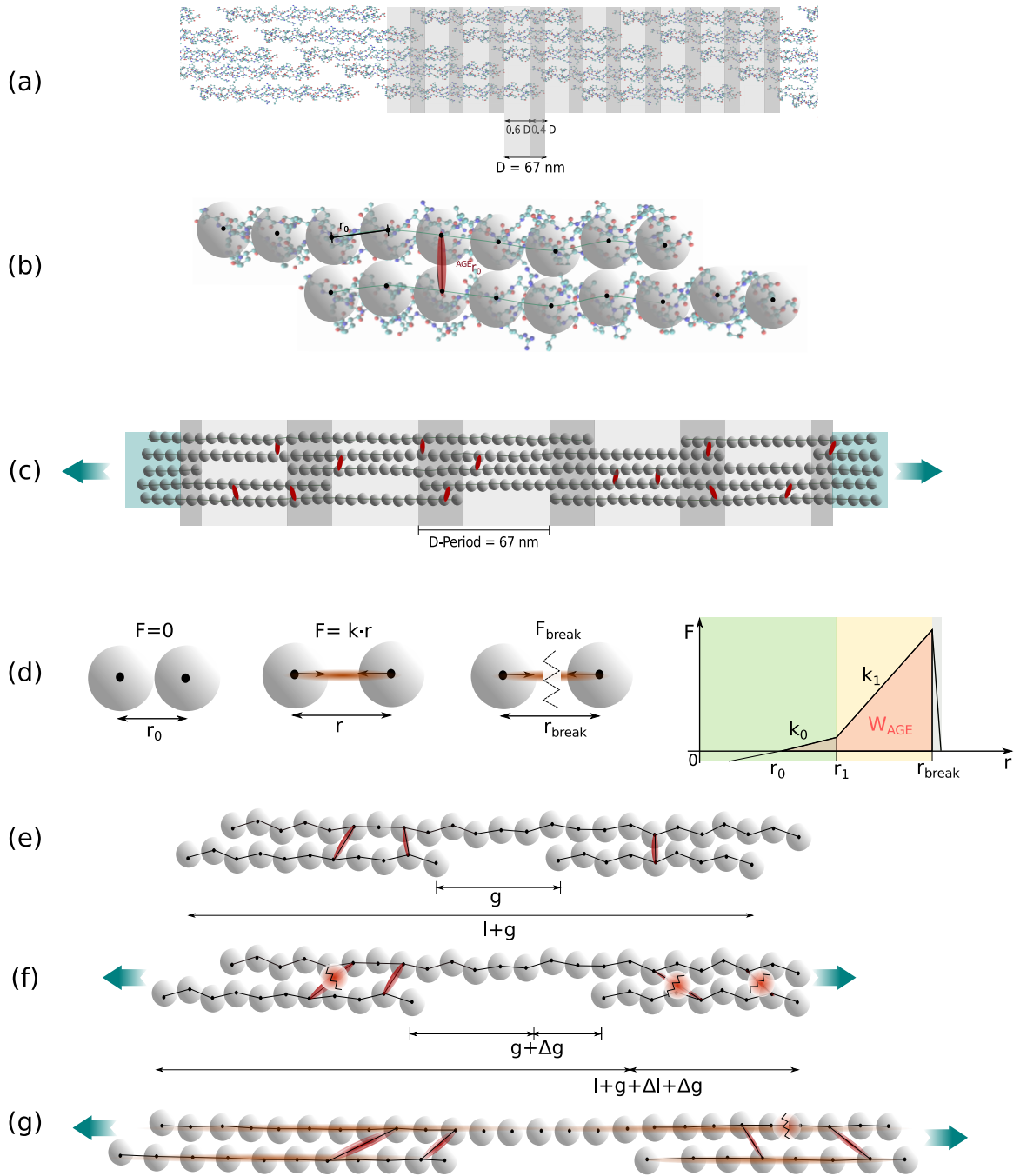

**Figure 1:** Schematic overview of model implementation and evaluation of deformation mechanisms. (a) Hodge-Petruska Model for collagen fibril: Displaying characteristic banding pattern with gap and overlap zones of TC molecules. (b) Coarse grained molecular dynamics model: The mechanical behavior of the TC molecules is represented by a string of particles mimicking the mechanical response that has been extracted from full scale simulations (Depalle et al., 2015). (c) Schematic representation of our implementation of a representative collagen fibril geometry: 5 gap and overlap zones; AGEs cross-links were randomly inserted between TC molecules (red) with different densities; fibril is strengthened at the ends (blue area) to guarantee smooth force transmission. (d) Definition of bond interactions (collagen bonds in TC molecules and AGEs cross-links): trilinear bond behavior, where the force depends on the distance $r$ between two particles. The varied parameters in our simulations are $k_1$ and $r_{break}$; the loading energy capacity of a single bond, *e.g.*, $W_{AGE}$ indicated by orange area. (e) Schematic representation of original configuration between two staggered TC molecules before tensile testing is started. (f) Sliding between TC molecules during tensile testing: AGEs cross-links rupture, the sliding of the TC molecules is responsible for energy dissipation in the collagen fibril, and the gap length $g$ is increased by $\Delta g$. (g) Stretching of the TC molecules during tensile testing: Load transmission to the TC molecules, AGEs withstand the force, eventually fracture of TC molecules.

Parameters used in coarse-grained molecular dynamics mesoscale model of collagen fibrils. Reference values for glucosepane are indicated in grey.

| Components | Parameters | Value |
|---|---|---|
| AGEs Cross-links | Equilibrium particle distance ($r_0$, Å) | 18.52 |
| | Critical hyperelastic distance ($r_1$, Å) | 22.72 |
| | Bond breaking distance ($r_{break}$, Å) | 25 - 27 - 30 - 33 - 35 - 40 - 50 - 60 |
| | Glucosepane: Bond breaking distance ($r_{break}$, Å) | 31.72 |
| | Tensile stiffness parameter ($k_0$, kcal, mol$^{-1}$Å$^{-2}$) | 0.1 |
| | Tensile stiffness parameter ($k_1$, kcal, mol$^{-1}$Å$^{-2}$) | 6 - 8 - 12 - 16 - 24 |
| | Glucosepane: Tensile stiffness parameter ($k_1$, kcal, mol$^{-1}$Å$^{-2}$)) | 8.00 |

of particles of two TC-molecules. Details about the exact insertion procedure of the cross-links are provided in Kamml et al. (2023). The exact composition, locations and numbers of AGEs in physiological collagen fibrils remain generally unknown. Exceptions are limited to studies of few individual types of AGEs that determined their location (Gautieri et al., 2014). For instance, glucosepane, the most prominent AGE, reaches levels of 2000 pmol/mol in collagen tissue of 90 years old patients, and ~ 4500 pmol/mol in diabetic patients, which corresponds to one AGE every 5 molecules and 2 molecules, respectively (Sell et al., 2005). However, many other types of AGEs, which have not been quantified so far, co-exist in collagen fibrils, and, hence, the exact numbers of cumulative AGEs cross-links is unknown. To explore the effects of AGEs cross-link density on the mechanical performance of the collagen fibril, and to account for their cumulative presence, we vary the inserted density and model cases with 0, 0.50, 1, 2, 5, 10, and 40 AGEs cross-links per TC molecule.

## 2.2. AGEs cross-link parameters

The total energy of the coarse-grained model is the sum of all pairwise particle interactions, as described by:

$$E_{total} = E_{bond} + E_{angle} + E_{inter}$$
$$= \sum_{bond} \Phi_{bond}(r) + \sum_{angle} \Phi_{angle}(\phi) + \sum_{inter} \Phi_{inter}(r) \quad (1)$$

where $E_{bond}$ is the bond energy due to stretching, $E_{angle}$ the dihedral bond interactions energy due to bending (for further description, see Appendix 8.1) , and $E_{inter}$ the pairwise interaction energy due to molecular interactions such as Van-der-Waals forces. Specifically, the bond energy represents the particle interactions through fixed interactions, as they occur within TC-molecules, but also between them as cross-links. In our model, there are two different types of bonds, which contribute to the bond energy via

$$\sum_{bond} \Phi_{bond}(r) = \sum_{collagen} \Phi_{collagen}(r) + \sum_{AGE} \Phi_{AGE}(r) \quad (2)$$

where $\Phi_{collagen}$ is the bond potential between particles within any given TC-molecule chain, and $\Phi_{AGE}$ is the bond potential of AGEs cross-links, linking particles between two

adjacent TC-molecules. The AGE cross-link force between two TC-molecules is then given by

$$F_{AGE}(r) = -\frac{\partial \Phi_{AGE}(r)}{\partial r} , \quad (3)$$

where $r$ is the radial distance. In our model, AGEs forces are modeled as trilinear springs as follows

$$F_{AGE}(r) = \begin{cases} -k_0(r - r_0) & \text{if } r < r_1 \\ -k_1(r - r_0) & \text{if } r_1 \leq r < r_{break} \\ z \cdot k_1(r - r_0) & \text{if } r_{break} \leq r < r_{break} + a \\ 0 & \text{if } r \geq r_{break} + a \end{cases} \quad (4)$$

where $r_0$ is the equilibrium distance between two linked particles, $k_t^{(0)}$ and $k_t^{(1)}$ are spring constants of the deformation, and $a$ is defined as $a = z \cdot (r_{break} - r_1)$. The chosen tri-linearity accounts for breaking of the bond, which occurs at $r_{break}$, and a gradual transition to its broken state via a regularization factor $z$ (Kamml et al., 2023). All other bond forces in our model follow the same trilinear behavior but with different parameters, as detailed in Appendix 8.1.

We perform a parameter study and vary strength and stiffness of AGEs cross-links. As a reference, we use the approximated values of glucosepane obtained from full-scale simulations with a reactive force field (Crippa, 2013; Kamml et al., 2023). We then vary stiffness and bond breaking distance (the inter-particle distance at which AGEs cross-links break), since these two values define the loading energy capacity of AGEs cross-links, meaning how much force it can withstand while stretching. The full range of applied AGEs parameters are displayed in Tab. 1.

## 2.3. Simulations

The simulations of the destructive tensile tests are performed in *LAMMPS* (Plimpton, 1995). The same procedure is applied as in our previous study (Kamml et al., 2023) using steered-molecular dynamics at a constant velocity of 0.0001 Å/fs (= 10m/s) with a time step of $\Delta t = 1$ fs. We verified in our previous study that lower strain rates do not lead to qualitative differences in simulation results (Kamml et al., 2023). The two ends of the fibril are constrained and moved apart during the tensile tests, and the required force is calculated and converted into engineering stress.

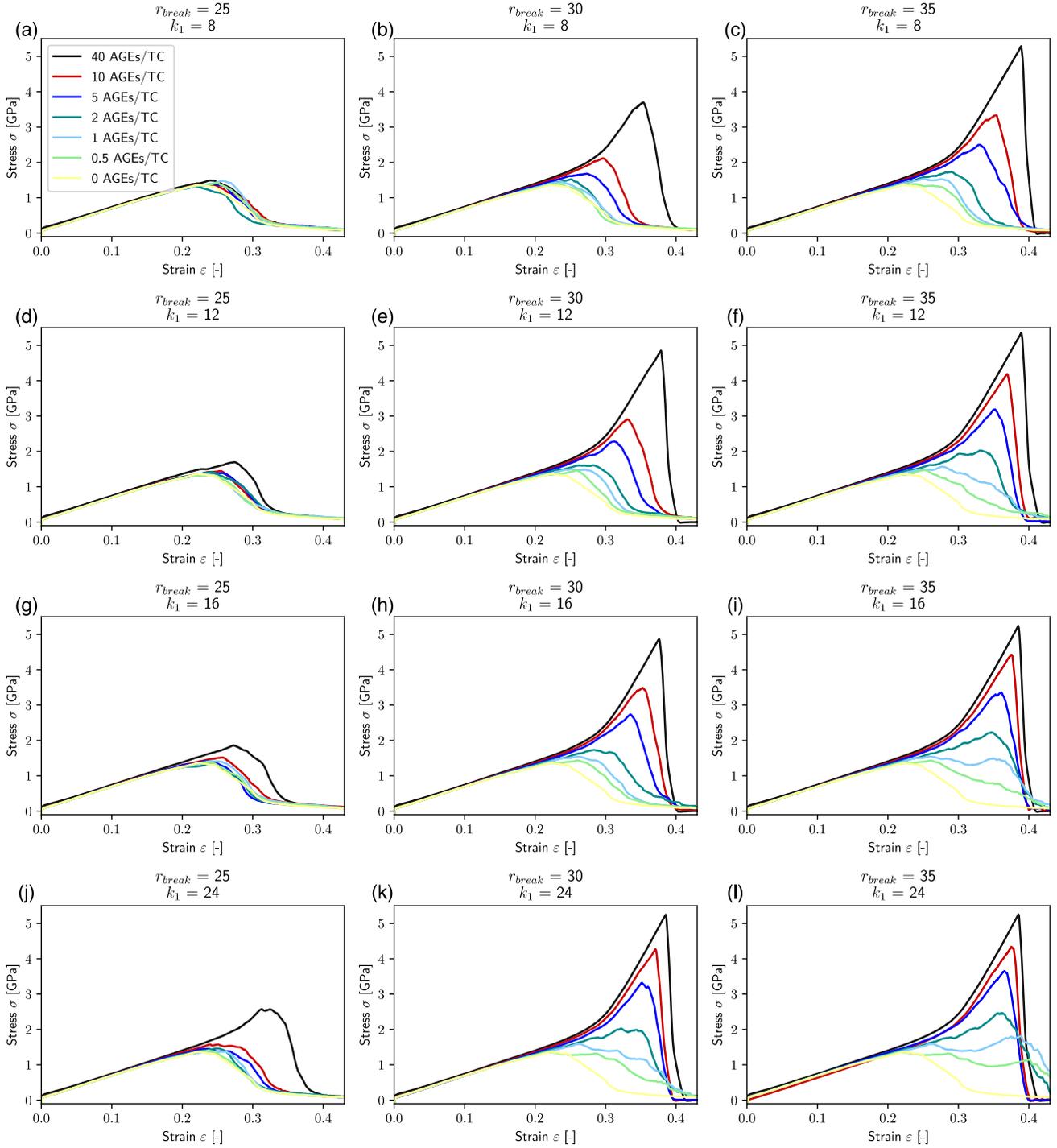

**Figure 2:** Stress-strain curves of destructive tensile tests of representative collagen fibril with different AGEs cross-link densities and varying AGEs tensile stiffness $k_1$ and critical breaking length $r_{break}$. The effect of glucosepane corresponds approximately to (b).

## 3. Results

### 3.1. Deformation and strength behavior

In our simulations, we vary the AGEs cross-links stiffness $k_1$ and the bond breaking distance $r_{break}$, which is the inter-particle distance at which AGEs cross-links break. We investigate the effect of changing these parameters on the mechanical properties of collagen fibrils with different AGEs densities (see Fig. 2). Our results show that various aspects of the macroscopic fibril behavior change significantly with changing stiffness and bond breaking length of the cross-links. Generally, the stress-strain curves show a two-phase behavior as observed in previous works (Kamml et al., 2023): at low strains $\varepsilon < \varepsilon_0 \approx 0.15$, where $\varepsilon_0$ defines the limit of the linear elastic regime, the fibril presents a linear deformation behavior. This initial linear behavior is independent of the cross-link parameters and density, *i.e.* the fibril stiffness is the same for all cases. Beyond the limit of $\varepsilon_0$, the fibril shows different mechanical behavior depending on stiffness $k_1$, breaking distance $r_{break}$ and cross-link density $N_{AGE}$. At low $N_{AGE}$ (0-1 AGEs/TC) and with low $k_1 < 24$ and $r_{break} \leq 30$ (see Fig. 2a, b, c, d, e, f, g, h, k, j), we observe a softening mechanism in the fibrils, where the stiffness reduces continuously until it does not have any remaining load bearing capacity, and hence failed. With larger $k_1$ and $r_{break}$ (*e.g.*, Fig. 2i, l) at the same $N_{AGE}$ (*i.e.* 0-1 AGEs/TC) the behavior is slightly different, where the stiffness only reduces temporarily, but then increases further to finally reach the maximum stress before failing somewhat more abruptly. This suggests that this type of AGEs may activate a different deformation mechanism of the fibril. Finally, fibrils with high AGEs densities, *i.e.* $N_{AGE} = 2 - 40$ AGEs/TC, show a very different post-$\varepsilon_0$ behaviour that is strongly dependent on $k_1$ and $r_{break}$ of the AGEs (see Fig. 2 & 4). When the AGEs are soft and/or weak (*i.e.* $k_1 < 24$ ; $r_{break} < 30$), the fibril shows a failure behavior with a smooth softening mechanisms, whereas stiff and strong AGEs may lead to fibrillar stiffening at $\varepsilon > \varepsilon_0$ depending on $N_{AGE}$.

These results show clearly that the AGEs properties, specifically their stiffness and critical failure length, have direct effects on the mechanical properties of the collagen fibril, and the extent of these effects depends on $N_{AGE}$. For example, the peak stress $\sigma_{peak}$ increases with increasing $k_1$, $r_{break}$ and $N_{AGE}$, see Fig. 2 and Fig. 3. Furthermore, $\sigma_{peak}$ presents saturation with $k_1$ and $r_{break}$, where further increases in $k_1$ and $r_{break}$ do not lead to higher $\sigma_{peak}$, *i.e.* $\sigma_{peak} \leq \sigma_{peak}^{sat}$. This saturation limit depends on the combination of $k_1$ and $r_{break}$. For example at $N_{AGE} = 2AGEs/TC$ (see Fig. 3a), the saturation is reached at $r_{break} \approx 45$ at $k_1 \approx 5.0$, but with increasing $k_1 \approx 16$, $\sigma_{peak}^{sat}$ is already reached at $r_{break} \approx 35$. Similar trends can be observed for other AGEs densities (see Fig. 3). However, we note that, first, $\sigma_{peak}^{sat}$ depends strongly on $N_{AGE}$ and increases for higher $N_{AGE}$ (see Fig. 3), and second, the saturation limit moves to lower $k_1$ and $r_{break}$ values.

The appearance of the stiffening regime with changing parameters is important, since this demonstrates a change in mechanical behavior of the collagen fibril. To characterize the onset of the stiffening regime, to evaluate its effect on $\sigma_{peak}$, and to analyze the AGEs properties causing it, we computed the elastic-to-peak stress difference $\Delta\sigma = \sigma_{peak} - \sigma_0$, with $\sigma_0 = \sigma(\varepsilon_0)$. Accordingly, $\Delta\sigma \geq 0$ indicates stiffening of the fibril, and $\Delta\sigma \approx 0$ corresponds to a fibril that does not stiffen before failure. Furthermore, a larger $\Delta\sigma$ indicates more pronounced stiffening, and since the linear regime is independent of the AGEs density, this also means that such a fibril presents a larger $\sigma_{peak}$ (compare Fig. 3 with Fig. 4). In general, we only observe significant stiffening for $r_{break} > 25$Å and for fibrils with $N_{AGE} \geq 2AGEs/TC$ (see Fig. 4).

In our previous work (Kamml et al., 2023), we showed that the stiffening of collagen fibrils with high AGEs density occurs because a high number of AGEs leads to small average forces within the AGEs and hence small AGEs deformation, which results in large deformations of the TC molecules. In other words, high AGEs density causes force transmission to be predominantly through the TC molecules (see Appendix 8.2), and hence fibrillar stiffness approaches the stiffness of the TC molecules. Whether this transition to TC-deformation-governed state occurs is the result of the relative stiffness between the AGEs and the TC molecules, which explains the above-discussed effect of $k_1$ on the stiffening limit (see Fig. 4). However, relative-stiffness solely is not sufficient to explain the link between the AGEs properties and their effect on the fibril behavior. If the AGEs are weak, fibrillar stiffening cannot occur because the AGEs break before they can activate the TC-deformation-governed state. Hence, a criterion combining both AGEs properties, $k_1$ and $r_{break}$, is required to describe the stiffening limit of collagen fibrils with different types of AGEs. Here, we suggest that this criterion is expressed in terms of energy capacity, which is defined as

$$\begin{aligned} W(k_0, k_1, r_0, r_1, r_{break}) &= k_0/2 \left(r_1 - r_0\right)^2 \\ &+ k_0 \left(r_1 - r_0\right)\left(r_{break} - r_1\right) \\ &+ k_1/2 \left(r_{break} - r_1\right)^2 \, , \end{aligned} \quad (5)$$

as also shown in Fig. 1d, and which is used to compute the loading energy capacity $W_{TC}$ and $W_{AGE}$ of the TC molecules and AGEs cross-links, respectively.

We argue that the balance between the energy capacity of a TC molecule and its attached cross-links, as expressed by

$$W_{TC} = N_{AGE} \cdot W_{AGE} \quad (6)$$

provides a reasonable criterion to describe the occurrence of collagen fibrillar stiffening (see the grey line in Fig. 4a-d). We observe that this balance is consistent with the onset of the stiffening regime. Specifically, for $W_{TC} < N_{AGE} \cdot W_{AGE}$ the TC-deformation-governed regime is activated. This observation suggests that the macroscopically observed changes are a direct result of modifications to the deformation mechanism at the nanoscale as caused by the presence of AGEs cross-links, which is best described in terms of relative loading energy capacity. It can be expected that such changes also effect energy absorption in the collagen fibril and, hence, their failure mechanism.

### 3.2. Failure mechanisms, energy dissipation and toughness

In Sec. 3.1, we showed that macroscopic changes in the fibrillar stiffness and strength are the result of different deformation mechanisms at the inter fibrillar level. Similar reasoning should also apply to the failure mechanism of the collagen fibril. It has been suggested that the deformation and failure mechanism is directly affected by various types of energy absorption in the collagen fibril, which are more or less dominant depending on the AGEs density (Acevedo et al., 2018b,a; Gautieri et al., 2017; Svensson et al., 2013; Fessel et al., 2014; Zimmermann et al., 2011). The two most important mechanisms for failure are stretching of the TC molecules, and energy dissipation through sliding between TC molecules.

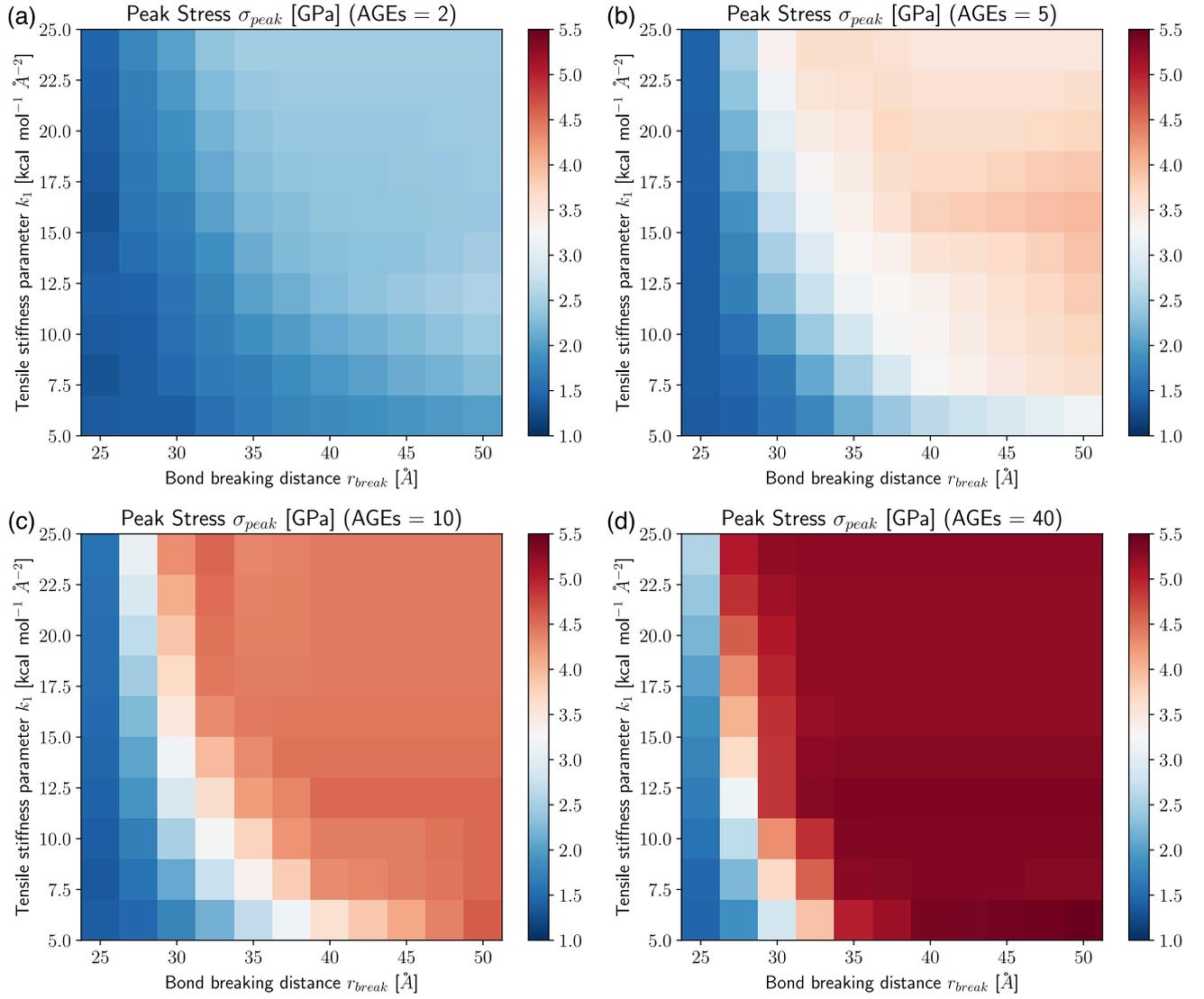

**Figure 3:** Peak stress $\sigma_{peak}$ of collagen fibril for varying tensile stiffness $k_1$ and bond breaking distance $r_{break}$ of AGEs cross-links at (a-d) AGEs densities of 2, 5, 10, 40 AGEs/TC, respectively.

We compute the various contributions to the fibrillar energy absorption to determine their overall implication in the failure process of the collagen fibril. The energy being absorbed in stretching of the TC molecules at a given global strain level $\varepsilon$ is given by

$$E_{stretch}(\varepsilon) = \int_{\Delta r_{TC}(\varepsilon^*)}^{\Delta r_{TC}(\varepsilon)} F_{stretch}\, \mathrm{d}\Delta r_{TC}\ , \qquad (7)$$

where $F_{stretch}$ is the average force within TC bonds, directly computed in the simulations, and $\Delta r_{TC}$ is the average change in length per collagen bond with respect to the equilibrium bond length ($\Delta r_{TC} = r_{TC} - r_0$) at a specific global strain level $\varepsilon$. We compute energy absorption by TC stretching starting from $\varepsilon^*$, where fibrillar strain is equal to the average strain in the collagen fibril ($\varepsilon = \varepsilon_{TC}$), which is approximately the transition point between TC unfolding and TC stretching-governed deformation mechanism.

Further, we estimate the cumulative energy dissipated through sliding using

$$E_{slide}(\varepsilon) = \int_{\Delta g(\varepsilon^*)}^{\Delta g(\varepsilon)} F_{slide}\, \mathrm{d}\Delta g\ , \qquad (8)$$

where $\Delta g$ is the change in gap size (see Fig. 1e-f) and $F_{slide}$ is the force due to friction of TC molecules. We note that this is a heuristic measure for sliding dissipation, which we use for simplicity. Specifically, the force $F_{slide}$ is the average force per TC molecule within the fibril that is not arising from bond forces of collagen bonds or cross-links and, hence, it is an estimate calculated from the globally applied tensile force $F$ on every TC molecules reduced by the forces due to stretching of the molecule. The change in gap size $\Delta g$ was obtained from the global collagen fibril strain $\varepsilon$ by approximating it with the average change in bond and gap length as

$$\varepsilon = \frac{\Delta g + n_{TC} \cdot \langle (r_{TC} - r_0) \rangle}{\langle g \rangle + n_{TC} \cdot r_0} , \qquad (9)$$

where $n_{TC}$ is the number of bonds per TC molecule, $\langle g \rangle$ the average initial gap length, and $\langle (r_{TC} - r_0) \rangle$ the average change in length of the collagen bonds (see Fig. 1e-g).

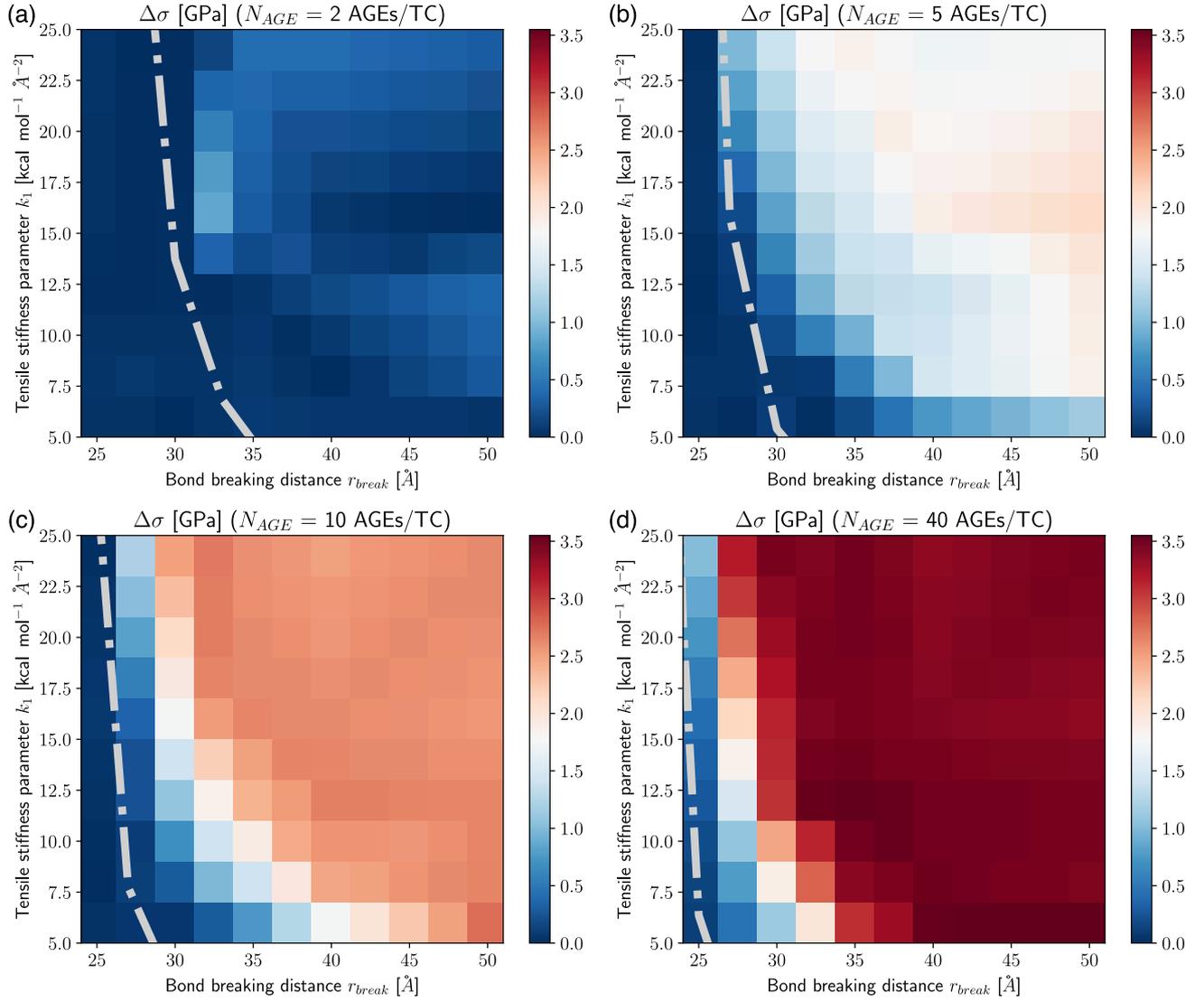

**Figure 4:** Elastic to peak stress change $\Delta\sigma$ of collagen fibril for varying tensile stiffness $k_1$ and bond breaking distance $r_{break}$ of AGEs cross-links at (a-d) AGEs densities of 2, 5, 10, 40 AGEs/TC, respectively. Dash-dotted line indicates theoretical limit for the existence of the stiffening regime, following Eq. 6.

First, we consider the effect of AGEs cross-link properties on the fibrillar energy absorption mechanisms. Specifically, we compute the energy absorption that occurs until failure initiates using Eq. 7 and 8 with $\varepsilon = \varepsilon_{peak}$, where $\sigma_{peak} = \sigma(\varepsilon_{peak})$. We normalize the respective energies with $E^0_{slide}$ and $E^0_{stretch}$, the energies of a collagen fibril with an AGEs density of $N_{AGE} = 0$, as it provides an AGEs-parameter-free reference value, and is expected to be the most extreme sliding case. We observe that $E_{slide}$ and $E_{stretch}$ are strongly dependent on both $r_{break}$ and $k_1$ (see Fig. 6a-d and 5a-d). Following the arguments presented in Sec. 3.1, we suggest that the most important mechanical property of the AGEs cross-link is the loading energy capacity $W_{AGE}$, which combines $r_{break}$ and $k_1$. If $E_{slide}$ and $E_{stretch}$ are reported with respect to $W_{AGE}$, the energy absorption is well described and can be directly linked to the AGEs property and the AGEs density (see Fig. 6e-h and 5e-h). Nevertheless, we note some limited scattering of the data, in particular in the sliding energy at low $N_{AGE}$ (e.g., Fig. 5e), which is likely due to the used heuristic measure that estimates the forces on the TC molecule due to sliding via force balance. Hence, tetrahedral forces, for instance, cannot be excluded from this estimate, and contribute indirectly to the sliding energy. These effects are stronger for cases with lower $N_{AGE}$ because overall forces in the fibril are lower. Despite this imprecision, the overall trend and importance of the AGEs energy capacity $W_{AGE}$ for the fibrillar behavior is evident, and allows us to describe and compare the effect of the stretching or sliding mechanisms on the collagen failure properties.

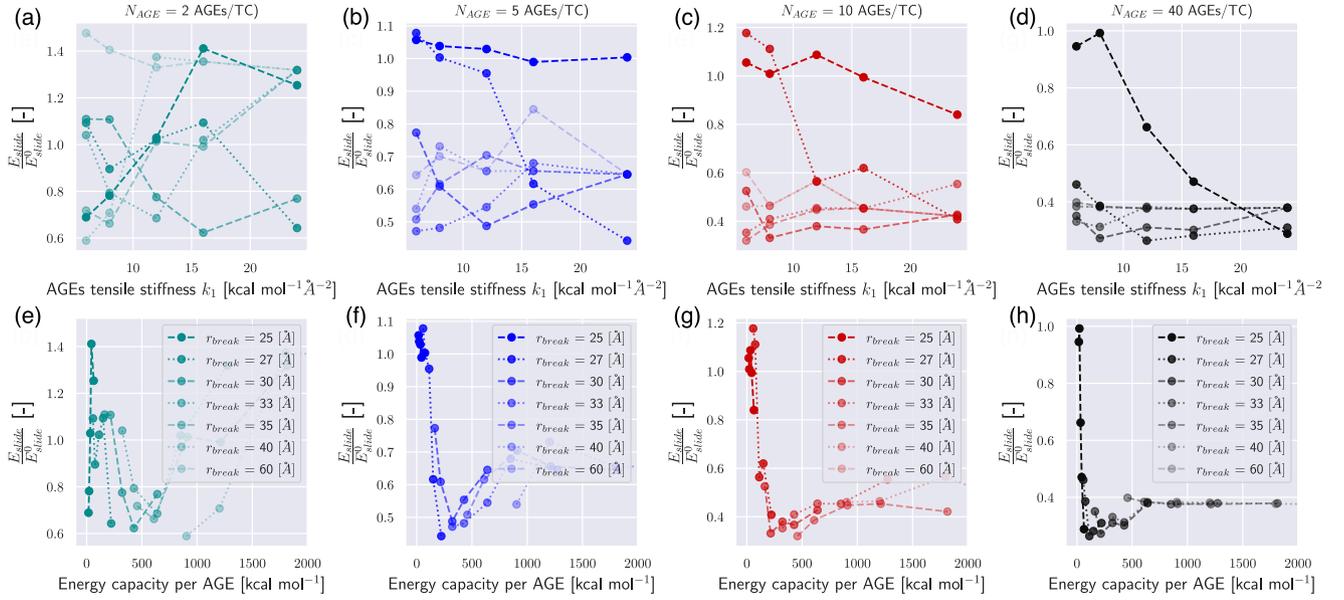

**Figure 5:** Sliding energy in collagen fibrils for different types of AGEs. (a-h) Sliding energy $E_{slide}$ computed up to global strain at peak stress $\varepsilon_{peak}$ following Eq. 8 normalized by the sliding energy $E_{slide}^0$ of a collagen fibril with 0 AGEs per TC molecule. (a-d) The normalized sliding energy is shown as function of AGEs tensile stiffness $k_1$ at varying $r_{break}$ for different AGEs densities, respectively. (e-h) The normalized sliding energy is shown as function of the AGEs loading energy capacity $W_{AGE}$ for different AGES densities, respectively.

For further evaluation of the influence of different $N_{AGE}$ and AGEs cross-link properties, we report the summarized and normalized data for stretching and sliding energy in Fig. 7. We see that sliding and stretching at lower levels of $W_{AGE}$ are dependent on both parameters (indicated by the dotted lines) up to a certain saturated value that is only dependent on $N_{AGE}$ (indicated by the dashed lines). The failure mechanism is either dominated by sliding or stretching and the relation between the two (compare Fig. 7a&b): We observe that with decreasing normalized sliding energy $E_{slide}$, normalized stretching energy $E_{stretch}$ increases, representing the relation of sliding and stretching as failure mechanism. At lower levels of $W_{AGE}$ and $N_{AGE}$ sliding is dominant whereas stretching is very limited. With both parameters increasing, sliding becomes less pronounced and stretching of the TC molecules is the more dominant mechanism. Both mechanisms then reach a saturation level dependent on the $N_{AGE}$ of the fibril. This shows that higher levels of $N_{AGE}$ and $W_{AGE}$ cause stiffening within the collagen through reduction of sliding between the TC molecules and by increasing their stretching. We note that since the sliding energy is expected to decrease with higher $N_{AGE}$, the reported $E_{slide}/E_{slide}^0$ should theoretically satisfy < 1. While most of our reported

$E_{slide}$ are consistent with this limit, some results for cases with $N_{AGE} = 2$ exceed it (see Fig. 7a). The likely cause is our heuristic approach to calculate $F_{slide}$, as discussed above, this may lead to relatively significant errors in low force situations such as fibrils with $N_{AGE} = 2$.

Since the induced stretching might eventually lead to fracture of the collagen bonds within the TC molecules, we evaluated the quantities of broken AGEs cross-links and broken collagen molecules in a collagen fibril at failure. We show the relative percentage of broken TC molecules (broken collagen bonds with respect to number of TC molecules) since broken TC molecules illustrate the extent of failure within the collagen fibril and imply a more abrupt failure. Therefore, the percentage of broken TC molecules reaches levels of more than 100%. We observe a pronounced decreasing trend of broken AGEs with increasing energy capacity of AGEs (see Fig. 8a). While the numbers of broken AGEs is dependent on $N_{AGE}$ for low energy levels, this trend vanishes for increasing energy levels. The decrease in broken AGEs cross-links with increasing load capacity is correlating with an increase in the number of broken TC bonds (see Fig. 8b). In fibrils, where their stretching has been fully activated due to the force transmission to the TC molecules, the saturated level of TC bonds to break is reached once their energy capacity is depleted. This results in fracture of the TC molecules, not the AGEs cross-links. At high energy capacities, the percentage of broken bonds in the TC molecule is highly dependent on $N_{AGE}$, due to the load transfer from AGEs to TC molecules (see Appendix 8.2). In summary, our results show that higher levels of $N_{AGE}$ and $W_{AGE}$ cause stiffening within the collagen fibril through reduction of sliding and enhancing of stretching of the TC molecules' backbone. This change in deformation mechanism is eventually causing fracture of the collagen bonds in the TC molecules by force transmission from AGEs cross-links to the TC molecules (see Appendix 8.2).

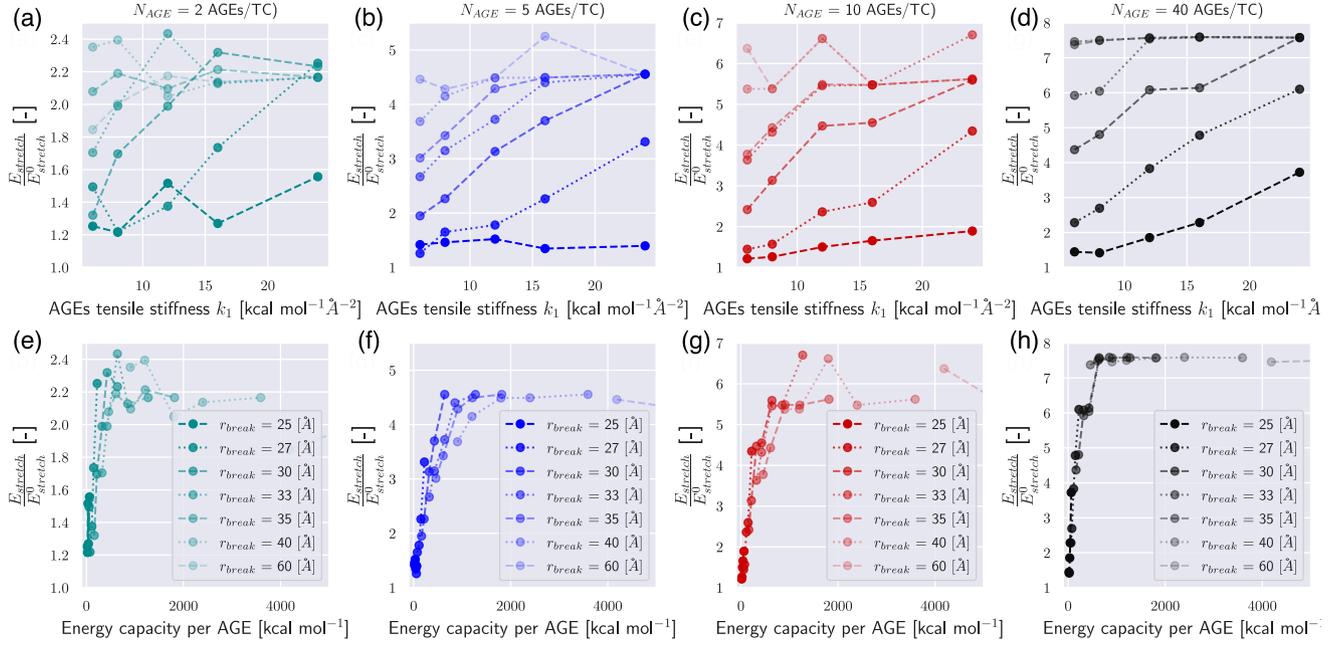

**Figure 6:** Stretching energy in collagen fibrils for different types of AGEs. (a-h) Stretching energy $E_{stretch}$ computed up to global strain at peak stress $\varepsilon_{peak}$ following Eq. 7 normalized by the stretching energy $E_{stretch}^0$ of a collagen fibril with 0 AGEs per TC molecule. (a-d) The normalized stretching energy is shown as function of AGEs tensile stiffness $k_1$ at varying $r_{break}$ for different AGEs densities, respectively. (e-h) The normalized stretching energy is shown as function of the AGEs loading energy capacity $W_{AGE}$ for different AGES densities, respectively.

On the macroscopic fibril level, we observe that fibrils that reach the stiffening regime, and hence present a high peak stress, also appear to fail in a significantly more brittle manner, as also noted in our previous study (Kamml et al., 2023). This behavior manifests itself as steep post-peak slopes in the stress-strain curves of the fibril (*e.g.*, see red and black curves in Fig. 2c,f,i,l). To estimate the fracture toughness of collagen fibrils, we consider the energy dissipated through sliding instead of the commonly applied measure of work to failure, which is not an appropriate measure for brittleness. Accordingly, our results show that the sliding energy, and hence collagen fibril toughness, decreases with $W_{AGE}$ and $N_{AGE}$, with the latter being the determining factor for the extent of this reduction. Since, such a deformation mechanism leads to more stretching of the TC molecules, their absorbed energy causes higher energy release at fibril rupture, resulting in a brittle failure of the entire collagen fibril. The $W_{AGE}$ accounts for different types of AGEs and we see, that especially at low levels of $N_{AGE}$, the type of AGE has a significant influence on the collagen deformation and failure mechanism by changing intrafibrilar deformation mechanisms.

## 4. Discussion

### 4.1. General limitations

With our model of a simplified collagen fibril, we can directly observe the effects of critical parameters on the mechanical response of the fibril during tensile testing. Nevertheless, in physiological conditions numerous other factors contribute to its varying mechanical behavior. For example, tissue hydration regulated via osmotic pressure is possibly a crucial factor of collagen mechanics in the body (Andriotis et al., 2019), which we did not account for. In general, AGEs cross-links only occur in mature tissue, in which non-enzymatic cross-links are also present. While this aspect was neglected in the present study, in our previous work (Kamml et al., 2023) we observed that enzymatic cross-links, which stabilize the collagen fibril, have an effect on the occurrence of the stiffening regime, but do not change the qualitative collagen fibril behavior.

Further, we note that our model considers AGEs distribution as random, since, with the exception of one specific AGE (glucosepane) (Gautieri et al., 2014), the exact location of AGEs along the TC molecule remains still unknown. In general, the lack of information about different types and, more importantly, quantities of AGEs cross-links present in specific tissues is a major problem. Currently, it is known that there are more than a dozen AGEs cross-link types that might occur in different combinations and quantities depending on the tissue type but their mechanical properties are mostly unknown. Given the limited available information, we used glucosepane ($k_1 = 8.00$ ; $r_{break} =$

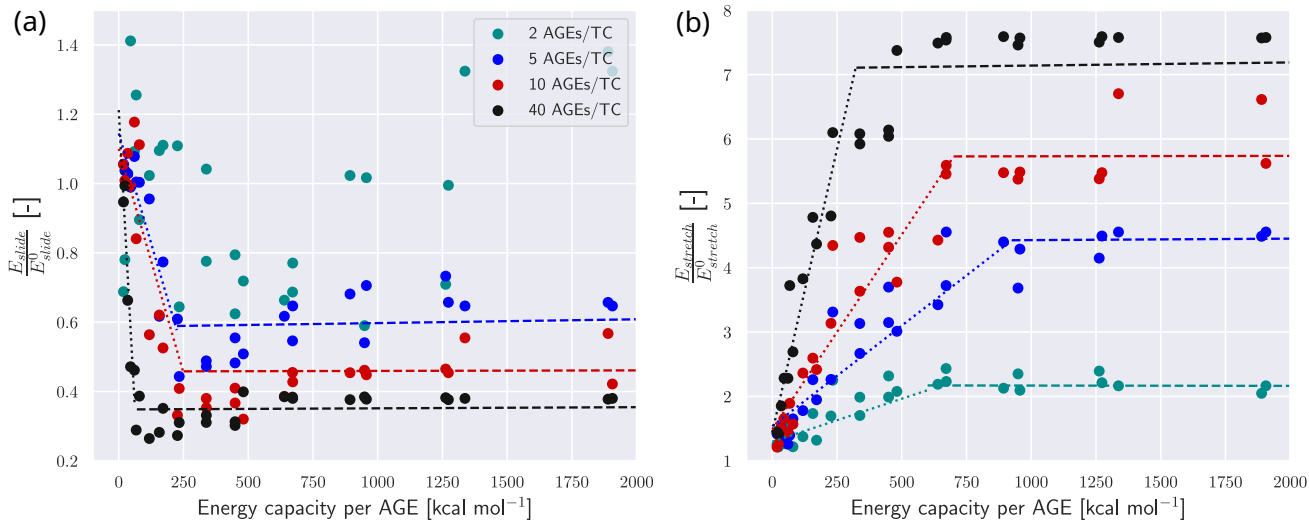

**Figure 7:** Energy absorption in collagen fibril. (a) Dissipated energy through sliding $E_{slide}$ normalized by $E_{slide}^0$, where $E_{slide}^0$ is the reference energy dissipated by sliding of fibrils with 0 AGEs. Dotted lines fitted to data up to minimum value of $E_{slide}/E_{slide}^0$. Dashed lines show the saturated value, fitted to data from minimum value of $E_{slide}/E_{slide}^0$. (b) Energy $E_{stretch}$ absorbed by stretching of the bonds within the TC molecules normalized by $E_{stretch}^0$, where $E_{stretch}^0$ is the reference energy absorbed via stretching of fibrils with 0 AGEs. Dotted lines linearly fitted to data where $E_{slide}/E_{slide}^0 < 0.75 \max(E_{slide}/E_{slide}^0)$. Dashed lines linearly fitted to data where $E_{slide}/E_{slide}^0 > 0.75 \max(E_{slide}/E_{slide}^0)$.

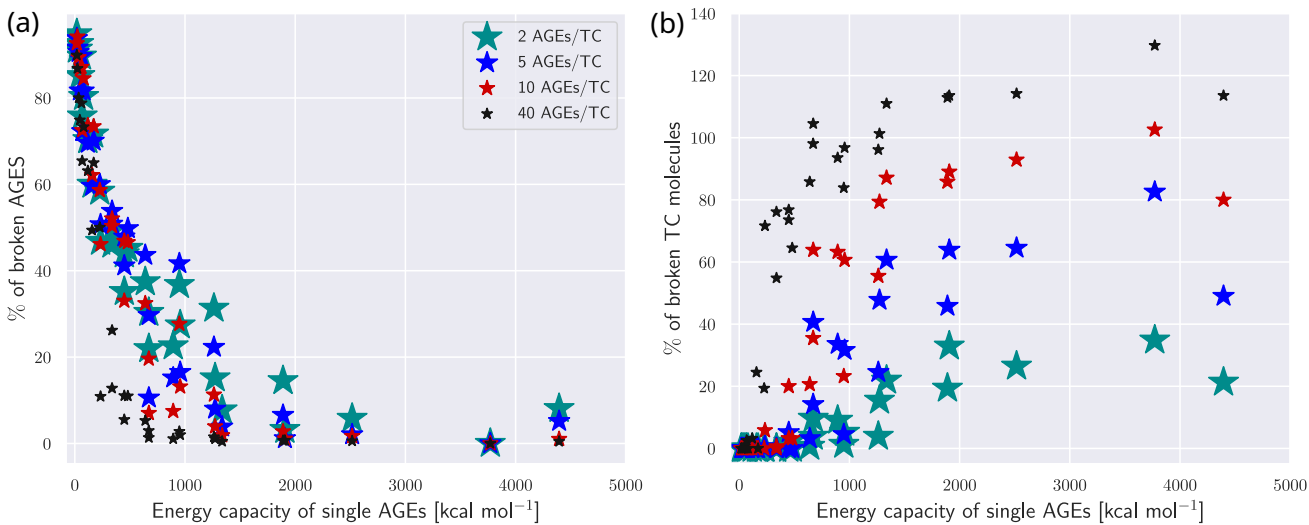

**Figure 8:** Broken bonds after failure of collagen fibril with AGEs cross-links. Percentage of broken (a) AGEs cross-links and (b) collagen bonds in TC molecule, shown as function of the loading energy capacity $W_{AGE}$ of AGEs. Broken bonds are counted after failure of the collagen fibril.

31.72, see Tab. 1), one of the most abundant cross-links, which appears to have relatively average mechanical properties, as a reference case and varied the applied mechanical properties of the cross-link with respect to it. Furthermore, the AGEs densities applied in our study exceeds the measured AGEs quantities for a single type of AGEs (Sell et al., 2005). However, the observed mechanical behavior is still likely to occur, since 1) there are more than one type of AGEs simultaneously present in collagen cross-links, and their effects add up, and 2) other types of AGEs with higher loading capacity of AGEs $W_{AGE}$, cause similar effects to the fibril at lower AGEs densities, as we have shown here.

Our results show that, apart from cross-link density $N_{AGE}$, the loading energy capacity $W_{AGE}$ of the AGEs cross-links, a nanoscale property, is the second contributing factor to changes in fibril deformation behavior. We found that the AGEs cross-link properties affects the level of AGEs density at which the collagen fibril experiences stiffening and how abrupt it fails, which depends on the energy absorption mechanism activated by the AGEs cross-links. Although our approach to estimate energy absorption via sliding is only heuristic, given the available information, it still allows us to qualitatively characterize the deformation and energy absorption mechanism, and it enabled us to demonstrate a causal link between an absence of frictional dissipation and fibril brittleness. Nevertheless, we note that an experimental validation of our results is needed to link our qualitative findings to quantitative observations in physiological tissue. Finally, we should also note that collagen tissue is not only built from collagen I, but also other types of collagen. This might additionally have influence on the material behavior at larger length scales and, hence, should be considered when applying our results to larger models.

In biological conditions, AGEs density and type, and hence their properties, are also dependent on their host tissue. Specifically, the number of present AGEs is dependent on the half life or turnover of the tissue, since AGEs form and accumulate during aging when tissue is exposed to increased glycation levels. This increase of AGEs on the nano-scale is likely to influence tissue mechanical properties on the macro-scale. The variability of collageneous tissues is wide and the unknown quantity of AGEs (cross-links) and types, apart from other contributing factors, prohibit currently any quantitative statements about collagen fibril mechanics. Nevertheless, our model reveals the general trends and the mechanisms caused by increased AGEs content and varying AGEs mechanical properties, and allows for interpretation of their influence on the macro-scale.

Another important factor in coarse-grained molecular dynamics is the choice of the applied force field. Our model of the particle interactions was obtained from literature (Depalle et al., 2015) and accounts for an implicit solvent, but hydration levels of tissue may also affect the mechanical response of the collagen fibril and its building components. While the simplifications associated to the force field definition are likely leading to quantitative differences, the qualitative mechanical deformation and fracture mechanics of the collagen fibril in in-silico simulations are expected to be consistent. Therefore, the obtained results can provide fundamental insight into the link between different AGEs cross-link properties and collagen mechanics, which is currently not possible in laboratory experiments due to size and resolution limitations.

### 4.2. Implications of the result

AGEs are a major concern when it comes to impaired properties of collageneous tissue, especially in patients with diabetes. The mechanisms leading to inferior mechanical performance in their presence remain poorly understood. In particular, knowledge about the types of AGEs that are present in various types of tissue, and in which quantities, is very limited. Consequently, it is also unknown how the properties of the AGEs affect collagen behavior and, for instance, which types of AGEs are the most critical.

Tissue behavior at the macro scale is highly dependent on the mechanics of the smaller scales. Our model of the collagen fibril aims at these issues at the nanoscale, where we perform tensile testing on the main building constituent of tissue, the collagen fibril. We account for the different types of AGEs cross-links by varying their mechanical properties (stiffness and bond breaking length) and reveal how these properties influence the deformation and failure behavior of the collagen fibril on the next larger scale. The original deformation and failure mechanism of collagen is sliding of the tropocollagen molecules within the fibril. Our results showed that the loading energy capacity $W_{AGE}$ of the AGEs cross-link combining both factors, stiffness $k_1$ and the breaking length $r_{break}$, is the second critical factor influencing the mechanical response of the collagen fibril apart from AGEs density, which has already been shown before (Kamml et al., 2023). When $W_{AGE}$ increases, the governing deformation and failure mechanism changes from sliding to stretching of the TC molecules, leading to fracture of the bonds within the TC molecules, which results in brittle failure. Hence, we conclude that it is not only AGEs density but also the mechanical properties of AGEs – specifically the loading energy capacity – that are responsible for changes in deformation and fracture behavior. This suggests that it is crucial 1) to determine the mechanical properties, such as $W_{AGE}$, of various types of AGEs, and 2) to focus on identifying the AGEs densities in priority for AGEs with high $W_{AGE}$ in a given tissue, as they are expected to have a stronger effect on the mechanical behavior of the collagen fibril.

The mechanical properties of collageneous tissues are highly dependent on the behavior of the collagen fibril, which is their main building constituent. Cross-link density and type are believed to be factors influencing collagen fibrils behavior and therefore, at the larger scale also the mechanical properties of tissue. For instance, brittleness in bone is most likely not only related to cross-link density but also cross-link type. Still, neither of them has been quantified in a laboratory study so far. It appears that the cross-link density is the governing factor determining the final strength of the fibril and the extent to which sliding between the TC molecules is reduced or stretching induced. Our results suggest that AGEs cross-links with increased $W_{AGE}$ have stronger influence on the reduction of sliding and toughness. Therefore, it would be important to quantify and evaluate the types of AGEs present in tissues, such as bone, to determine their effects on the impaired properties at the tissue level. The effect of different AGEs is important to understand the failure mechanism of collagen fibrils, and hence its consequences on the behavior of collageneous tissues, *e.g.* on the toughness of the bone. Our results support the perspective of Zimmermann et al. (2011) and Acevedo et al. (2018b,a), that AGEs cause bone brittleness by reducing sliding and thereby the natural energy dissipation mechanisms in bone. Since friction is limited within the collagen fibril due to this reduced sliding leading to decreased energy dissipation, the energy is absorbed by the TC molecules via stretching and the fibril appears to be more brittle at fracture. Additionally, the collagen fibril might not be the origin of fracture in bone with high AGEs content, since its stiffness and strength are increased. Instead, fracture might be initiated due to load transfer to the mineral component surrounding the fibrils Siegmund et al. (2008). The mineral in bone is known to be brittle and these mechanisms might contribute to the changes in fracture behavior. Further models including mineral should target these changes in bone collagen fibril mechanics.

Other tissues also show changes in mechanical behavior that correlate with the accumulation of AGEs. In tendons, for instance, an increase in failure stress (and strain) was observed. Specifically, on the fascicle and fibril level, the peak modulus was increased, as was also shown in our previous study (Kamml et al., 2023). Interestingly, Gautieri et al. (2017); Li et al. (2013); Svensson et al. (2018) suggest that AGEs reduce tissue elasticity in tendons by limiting fiber-fiber and fibril-fibril sliding, which is consistent with our findings, whereas Zellers et al. (2021) did not observe any statistically significant relationship between AGEs content and mechanical parameters, and claim that mechanical properties of tendons rather change due to collagen disorganisation than AGEs content. These observations need further investigations, since as stated, AGEs quantity and properties are highly dependent on the tissue. Another example are intervertebral discs, in which increased AGEs levels were shown to correlate with functional changes such as increased stiffness, increased torque range and increased torque failure (Wagner et al., 2006; Krishnamoorthy et al., 2018). Furthermore, in cartilage, tensile properties are changed and AGEs cause increased stiffness and strength, effects that we also observed, and decrease the failure length leading to brittleness (Verzijl et al., 2002; Chen et al., 2002), but it appears that cartilage network embrittlement at the nanoscale is responsible for increasing stiffness (Moshtagh et al., 2018), which still needs further investigation, since also other influences might contribute to changes in tissue mechanics.

Natural collageneous tissues are complex materials influenced by many factors, where AGEs are only one of them. Still, we can provide an insight into collagen fibril mechanics on the nano-scale and contribute to a fundamental understanding. Nevertheless, we lack consistent data of AGEs function, quantity and type (mechanical properties) in order to get further insight into collagen fibril mechanics. Additionally, since our study is numerical, experimental data of mechanical tests of collagen fibrils supporting the model implementation by calibration could contribute to a more consistent observation of collagen fibril mechanics in order to address the above mentioned limitations.

## 5. Conclusion

We performed a parameter study on an in-silico destructive tensile test of collagen fibrils at the nano-scale to analyze the effect of mechanical properties of AGEs cross-links, as they vary from AGEs type to type, on the collagen fibril behavior. We specifically focused on the deformation and fracture behavior of the collagen fibril and the associated energy absorption mechanisms. We found that apart from AGEs density, the other crucial factor for the stiffening of the collagen fibril is the loading energy capacity of the AGEs cross-links. Further, we showed that AGEs with higher loading energy capacity have a stronger impact on the fibrillar mechanics, leading to stiffening of the fibril and more brittle failure. We demonstrated that this effect is due to a change in the energy absorption mechanisms at smaller scales, where the presence of AGEs reduces inter-molecular sliding, which leads to less energy dissipation, and increases stretching of the TC molecules. Consequently, the fibril fails by fracture of the TC molecules, which is a low-toughness failure mechanism compared to failure by sliding within the fibril. These effects are generally more prominent in collagen fibrils with higher AGEs content, but the critical amount of AGEs cross-links decreases for AGEs types with higher loading energy capacity. Therefore, our results show that apart from the AGEs density, knowledge about the type of AGEs with their specific mechanical properties, is crucial for a better understanding of how the presence of AGEs impairs the behavior of collageneous tissue.

## 6. Acknowledgement

We are grateful to James L. Rosenberg (Utah), Ihsan Elnunu (Utah), Michal Hlobil (ETH) and Dr. Hajar Razi (ETH) for helpful discussions.

## 7. Funding

Research reported in this publication was supported by NIAMS of the National Institutes of Health under award number 1R21AR077881.

## CRediT authorship contribution statement

**Julia Kamml:** Formal analysis, Investigation, Visualization, Writing - Original Draft . **Claire Acevedo:** Conceptualization, Funding acquisition, Writing - Review & Editing . **David S. Kammer:** Conceptualization, Methodology, Funding acquisition, Writing - Review & Editing .

# 8. Appendix

## 8.1. General Force Field Modelling and Parameterization

For a comprehensive description of the modelling procedure and the force field definition, please refer to Kamml et al. (2023) and Depalle et al. (2015). Here we give a short overview of the parameters used for implementation.

In molecular dynamics the total potential energy $E_{total}$ of a system is defined by the sum of all contributing energies due to bonded and non-bonded interactions

$$
\begin{aligned}
E_{total} &= E_{bond} + E_{angle} + E_{inter} \\
&= \sum_{bond} \Phi_{bond}(r) + \sum_{angle} \Phi_{angle}(\phi) + \sum_{inter} \Phi_{inter}(r) \,,
\end{aligned}
\tag{10}
$$

where $E_{bond}$ is the bond energy due to stretching, $E_{angle}$ the dihedral bond interactions energy due to bending and $E_{inter}$ the pairwise interaction energy due to molecular interactions such as Van-der-Waals forces. Forces define the interactions between the particles and are calculated as the negative derivative of the potential energy, depending on the particle distance

$$
F = -\frac{\partial \Phi(r)}{\partial r}
\tag{11}
$$

or angle

$$
F = -\frac{\partial \Phi(\phi)}{\partial \phi}
\tag{12}
$$

For the definition of the bond interactions and the equations used for force calculation due to bonds, please refer to Sec. 2.2. The dihedral bond interactions due to the tropocollagen molecule geometry not being purely straight are modelled as

$$
F_{angle}(\phi) = -k_B(\phi - \phi_i) \cdot \phi \,,
\tag{13}
$$

with $\phi$ as the angle between three particles causing forces when not in equilibrium position $\phi_0$ and $k_B$ accounting for the bending stiffness.

The non-bonded interactions are modelled via a Lennard-Jones potential with a soft-core as follows

$$
F_{\text{non-bond}}(r) = \begin{cases} F_{LJ}(r) & \text{if } r \geq \lambda \sigma_{LJ} \\ F_{LJ}(\lambda \sigma_{LJ}) & \text{if } r < \lambda \sigma_{LJ} \end{cases}
\tag{14}
$$

where

$$
F_{LJ}(r) = \frac{1}{r}\left[48\epsilon_{LJ}\left(\frac{\sigma_{LJ}}{r}\right)^{12} - 24\epsilon_{LJ}\left(\frac{\sigma_{LJ}}{r}\right)^6\right].
\tag{15}
$$

The parameter $\epsilon_{LJ}$ is the well depth between two particles, $\sigma_{LJ}$ the distance at which the intermolecular potential between the two particles is zero and $\lambda$ the parameter to adjust the critical force associated to the soft core. We obtained the parameters of the interactions in collagen molecules and for the LJ potential for our coarse-grained model from full-scale simulations in literature and adjusted the approach by adding a soft-core to the LJ Potential (Depalle et al., 2015; Buehler, 2006a). Further, for the definitions of parameters of AGEs cross-links, we performed full-scale steered molecular dynamics tensile tests on glucosepane as a reference cross-link with a reactive force field. For parameters applied in our models, see Tab. 2.

**Table 2**
Parameters used in coarse-grained molecular dynamics mesoscale model of collagen fibrils (Depalle et al., 2015)

| Components | Parameters | Value |
|---|---|---|
| Collagen molecules | Equilibrium particle distance ($r_0$, Å) | 14.00 |
| | Critical hyperelastic distance ($r_1$, Å) | 18.20 |
| | Bond breaking distance ($r_{break}$, Å) | 21.00 |
| | Tensile stiffness parameter ($k_0$, kcal mol$^{-1}$ Å$^{-2}$) | 17.13 |
| | Tensile stiffness parameter ($k_1$, kcal mol$^{-1}$ Å$^{-2}$) | 97.66 |
| | Regularization factor ($z$, -) | 0.05 |
| | Equilibrium angle ($\phi_0$, degree) | 170.0 to 180.0 |
| | Bending stiffness parameter ($k_b$, kcal mol$^{-1}$ rad$^{-2}$) | 14.98 |
| | Dispersive parameter ($\varepsilon_{LJ}$, kcal mol$^{-1}$) | 6.87 |
| | Dispersive parameter ($\sigma_{LJ}$, Å) | 14.72 |
| | Soft core parameter ($\lambda$, -) | 0.9 |
| Particles at ends of TC molecules | *same parameters as collagen molecules except:* | |
| | Bond breaking distance ($r_{break}$, Å) | 70.00 |
| ECLs | Equilibrium particle distance ($r_0$, Å) | 18.52 |
| *Trivalent Cross-links* | Critical hyperelastic distance ($r_1$, Å) | 22.12 |
| | Bond breaking distance ($r_{break}$, Å) | 24.81 |
| | Tensile stiffness parameter ($k_0$, kcal, mol$^{-1}$ Å$^{-2}$) | 0.20 |
| | Tensile stiffness parameter ($k_1$, kcal, mol$^{-1}$ Å$^{-2}$) | 54.60 |
| | Mass of each mesoscale particle, atomic mass units | 1358.7 |

## 8.2. Additional results

In the following, we provide additional results to give a comprehensive overview of the evaluated data. During the stiffening of the collagen fibril with increased AGEs cross-link content, we observe a force transmission from AGEs cross-link to the backbone of the TC molecules: Average forces at $\sigma_{\text{peak}}$ in AGEs decrease with increasing $N_{AGE}$, increasing $k_1$ and $r_{break}$ (see Fig. 9), while average forces in bonds within the TC molecule increase (see Fig. 10. This is the force transmission, that leads to stiffening of the collagen fibril and changed energy absorption via molecular stretching rather than sliding.

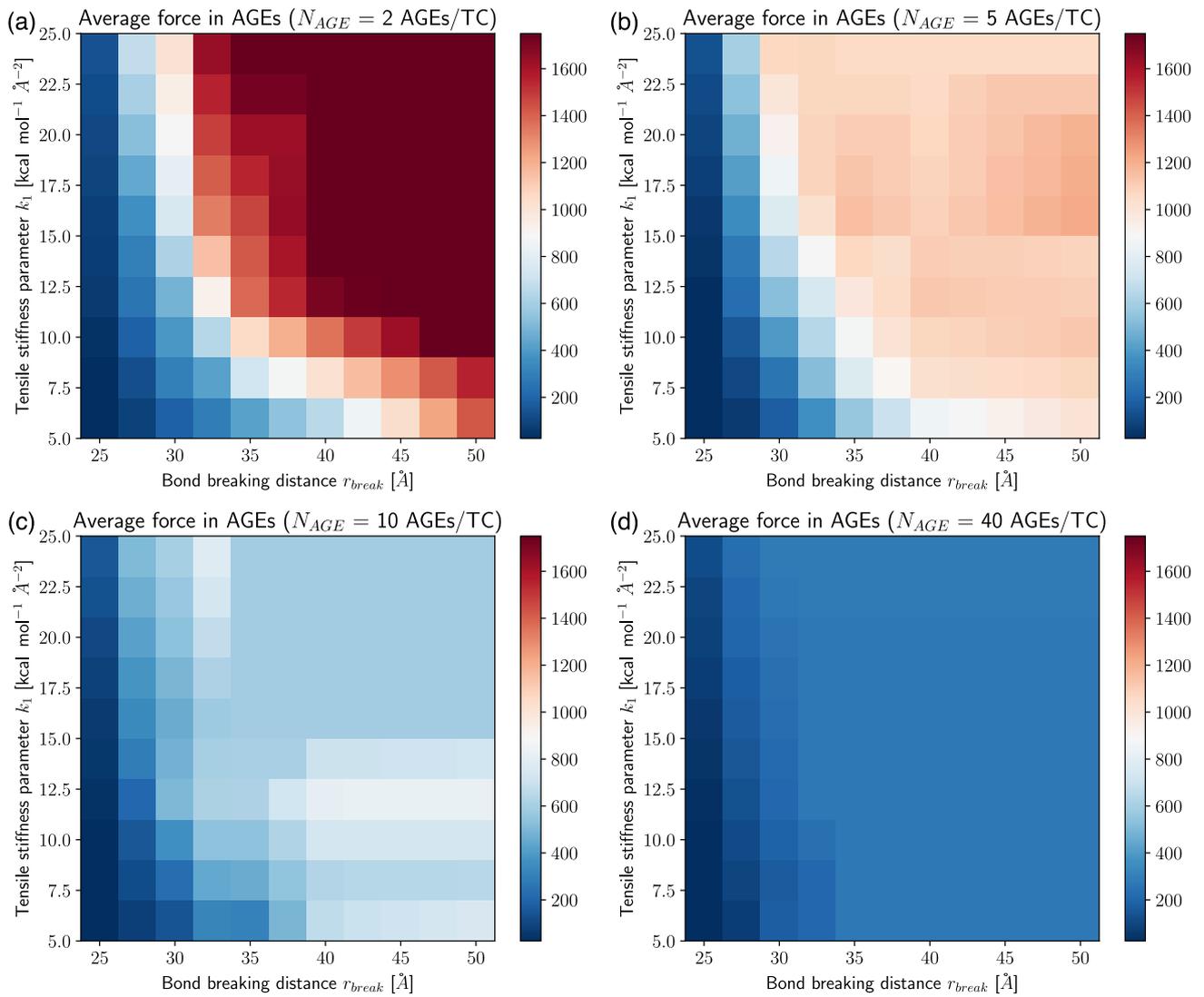

**Figure 9:** Average force $[(\frac{kcal}{mol})\text{Å}^{-1}]$ in AGEs cross-links at maximum stress $\sigma_{\text{peak}}$ at different AGEs densities and changing $k_1$ and $r_{break}$

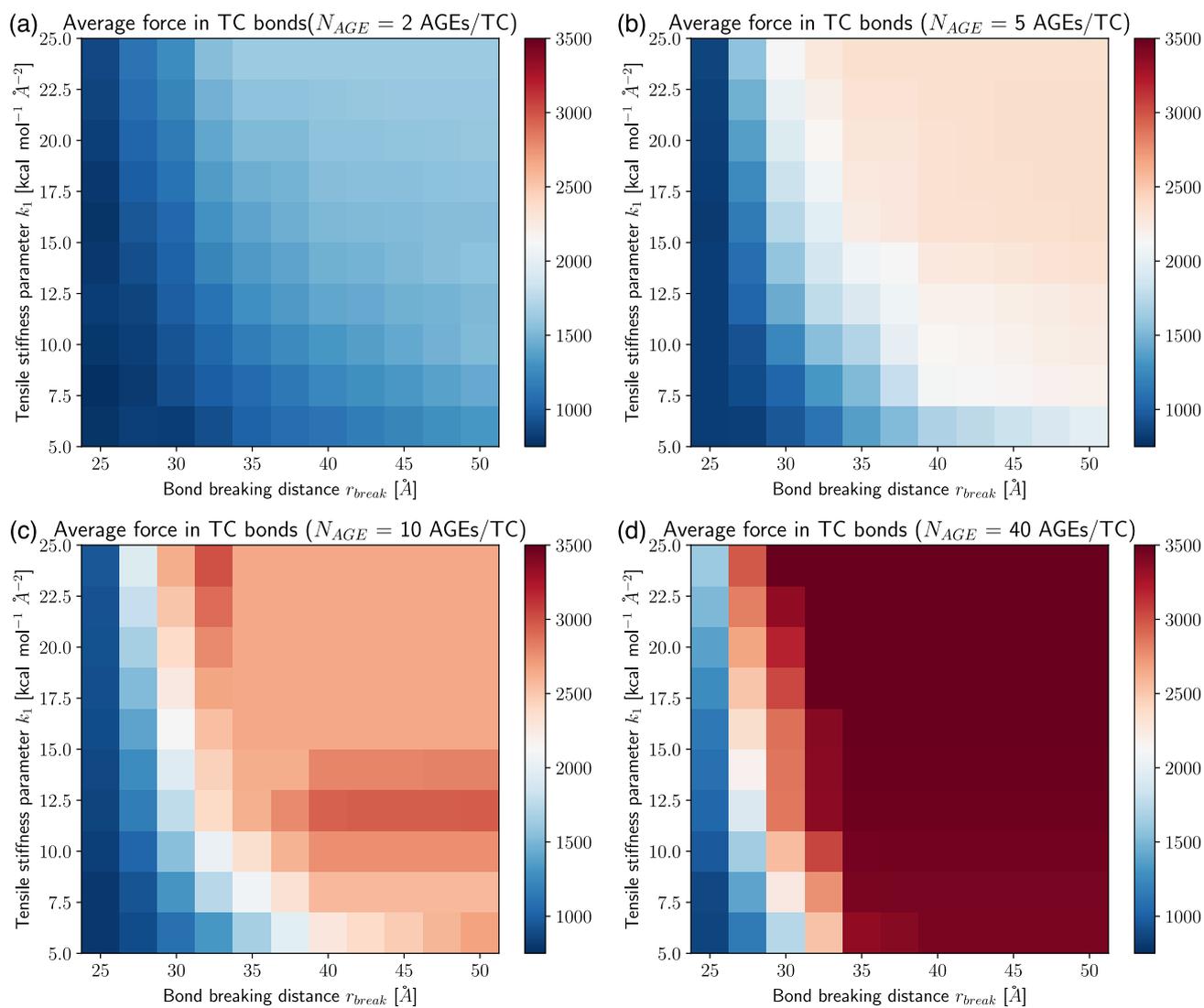

**Figure 10:** Average force $[(\frac{kcal}{mol})\mathring{\text{A}}^{-1}]$ in bonds within the TC molecule chains at maximum stress $\sigma_{\text{peak}}$ at different AGEs densities and changing $k_1$ and $r_{break}$